\documentclass[10pt,aps,prl,twocolumn,superscriptaddress]{revtex4}
\usepackage[linktocpage=true, colorlinks=true, urlcolor=blue, linkcolor=blue, citecolor=blue]{hyperref}
\usepackage{graphicx}
\usepackage{amsmath,amssymb}
\usepackage{color}
\usepackage{url}
\usepackage{hyperref}
\usepackage{cleveref}
\usepackage{xfrac}
\usepackage{upgreek}
\usepackage{epstopdf}

\graphicspath{{./Images/}}

\begin{document}

\title{Spatio-temporal filtering in laser Doppler holography for retinal blood flow imaging}

\author{L\'eo Puyo}
\affiliation{Corresponding author: gl.puyo@gmail.com}
\affiliation{Centre Hospitalier National d'Ophtalmologie des Quinze-Vingts, INSERM-DHOS CIC 1423. 28 rue de Charenton, 75012 Paris France}
\affiliation{Institut de la Vision-Sorbonne Universit\'es. 17 rue Moreau, 75012 Paris France}

\author{Michel Paques}
\affiliation{Centre Hospitalier National d'Ophtalmologie des Quinze-Vingts, INSERM-DHOS CIC 1423. 28 rue de Charenton, 75012 Paris France}
\affiliation{Institut de la Vision-Sorbonne Universit\'es. 17 rue Moreau, 75012 Paris France}

\author{Michael Atlan}

\affiliation{Institut Langevin. Centre National de la Recherche Scientifique (CNRS). Paris Sciences \& Lettres (PSL University). \'Ecole Sup\'erieure de Physique et de Chimie Industrielles (ESPCI Paris) - 1 rue Jussieu. 75005 Paris France}

\date{\today}

\begin{abstract}
Laser Doppler holography (LDH) is a full-field interferometric imaging technique recently applied in ophthalmology to measure blood flow, a parameter of high clinical interest. From the temporal fluctuations of digital holograms acquired at ultrafast frame rates, LDH reveals retinal and choroidal blood flow with a few milliseconds of temporal resolution. However LDH experiences difficulties to detect slower blood flow as it requires to work with low Doppler frequency shifts which are corrupted by eye motion. We here demonstrate the use of a spatio-temporal decomposition adapted from Doppler ultrasound that provides a basis appropriate to the discrimination of blood flow from eye motion. A singular value decomposition (SVD) can be used as a simple, robust, and efficient way to separate the Doppler fluctuations of blood flow from those of strong spatial coherence such as eye motion. We show that the SVD outperforms the conventional Fourier based filter to reveal slower blood flow, and dramatically improves the ability of LDH to reveal vessels of smaller size or with a pathologically reduced blood flow.
\end{abstract}

\maketitle

\section{Introduction}
Retinal motion caused by fixational eye movements and pulsatile swelling of the choroid prove challenging in ophthalmic imaging. Insufficient temporal resolution results in image blurring, or calls for complex registration schemes in the case of scanning methods. For interferometric imaging, eye motion additionally affects the backscattered light by the Doppler effect, which is especially problematic when the sought contrast is based on motion to measure blood flow or cellular activity. By now, hardware or software based methods have been developed to cope with the harmful effect of eye motion in many ophthalmic technologies. For example, in optical coherence tomography (OCT), it has been proposed to compensate eye motion using eye movements tracking with other modalities~\cite{Zawadzki2013, Baghaie2017}. In OCT-angiography (OCT-A), the scanning pattern can be modified and motion artifacts can be detected and corrected for~\cite{Chen2018, Kraus2014, Liu2014, Camino2016}. In Doppler-OCT, the bulk motion Doppler shift is also estimated and compensated for by various methods~\cite{Makita2006, Leitgeb2014}. Axial motion can also be digitally corrected in full-field OCT by using a phase gradient autofocus algorithm~\cite{Pfaffle2017}.

Laser Doppler holography (LDH) is a digital holographic method where blood flow is measured from the interference of coherent light backscattered by the eye with a reference beam~\cite{SimonuttiPaquesSahel2010, MagnainCastelBoucneau2014, Pellizzari2016}. The coherent gain brought by the reference beam allows to use ultrahigh camera frame rates to perform full-field measurement of the local Doppler broadening. The power Doppler calculated pixel-wise as the integral of the high-pass filtered Doppler power spectrum density (DPSD) reveals blood flow thanks to the larger Doppler broadening of light scattered by red cells. By using a sliding short-time window, the variations over cardiac cycles of retinal blood flow measured in power Doppler units can be measured with a few milliseconds of temporal resolution~\cite{Puyo2018}.
%
LDH is by nature very sensitive to eye motion as it affects all of the light backscattered by the eye which creates a strong contribution to the power Doppler. Global eye motion has for effect to decrease the blood flow contrast derived from the detection of light scattered by cells flowing in blood vessels. Intermittent motion especially hinders the measurement of blood flow variations. The pulsatile axial motion of the eye is especially challenging because it is partly synchronous to blood flow and occurs in the direction maximizing the Doppler effect. In our previous work, we imaged blood flow in large vessels revealed by frequencies in the tens of kHz range, while removing eye motion occurring in the range of a few kHz~\cite{Puyo2018}. By thresholding the power spectrum with a lower cutoff frequency of approximately 6-10 kHz, fast blood flow can be effectively isolated from eye motion. However when interested in slow blood flow or when in the presence of strong eye motion, the Fourier space does not allow to separate these contributions. Yet, the detection of blood flow in smaller or pathological vessels is of crucial clinical interest.
%
Interestingly, LDH shares the problem of slow flow detection with Doppler ultrasound imaging. The physical equations, biomarkers, and challenges to overcome are remarkably similar in optical and acoustic Doppler imaging, and one of these challenges is tissue motion. Ultrafast ultrasonic imaging, a technique based on the unfocused transmission of plane or diverging wave instead of the conventional focused emissions~\cite{TanterFink2014}, has allowed the parallel measurement of blood flow, increasing the acquisition frame rate by more than an order of magnitude. This subsequently allowed to use spatio-temporal analysis of the Doppler fluctuations to filter tissue motion~\cite{Demene2015, Alberti2017}, as it is also similarly done in x-ray computed tomography (CT)~\cite{Gao2011}, and magnetic resonance imaging (MRI)~\cite{Otazo2015}. Eigen-based clutter filters make use of the differences in spatio-temporal characteristics of tissue and blood motion for a better discrimination. In Doppler ultrasound, it was found that a singular value decomposition (SVD) which is a eigendecomposition of a non-square matrix, generates a basis where tissue and blood motion can be effectively separated. A similar method has also been implemented for full-field interferometric measurement to remove motion artifacts~\cite{Scholler2019}. We here report on the use of such type of spatio-temporal filter in the field of LDH.

\section{Method}
\begin{figure}[t!]
\centering
\includegraphics[width = 1\linewidth]{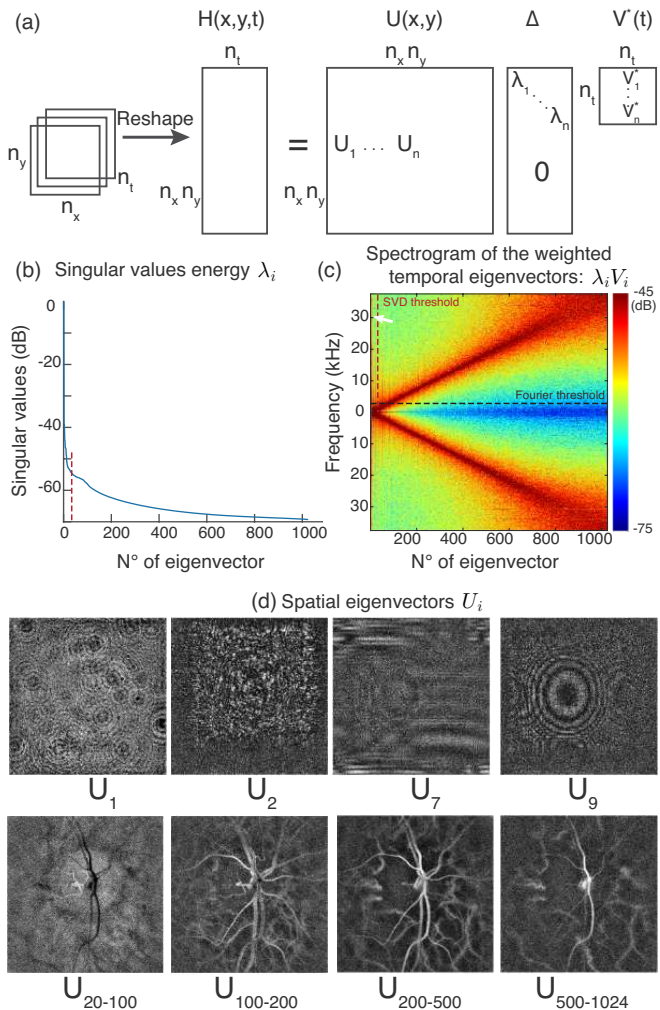}
\caption{Singular value decomposition (SVD). (a) The 3D matrix of complex-valued holograms $H(x,y,t)$ is reshaped into a 2D space-time matrix, and decomposed in a product of 3 matrices following Eq.~\ref{eq:eq_SVD}. $U$ and $V$ are the spatial and temporal eigenvectors, and $\Delta$ is the diagonal matrix of singular values $\lambda_i$. (b) Ordered singular values in dB. (c) Fourier transform magnitude of the temporal eigenvectors weighted by singular values, the arrow indicates high frequency clutter. (d) Individual or averaged spatial eigenvectors: the first vectors show clutter whereas vectors associated to singular values of lower energy reveal blood flow.
}
\label{fig_SVD_Processing}
\end{figure}

We use the fiber Mach-Zehnder LDH setup presented in~\cite{Puyo2018}. The light source used for the experiments is a 50 mW, single-mode laser diode at wavelength, 785nm (Thorlabs LP785-SAV50, VHG Wavelength-Stabilized SF Laser Diode, Internal Isolator). The power of the laser beam incident at the cornea was 1.5 mW of constant exposure, and the retina was imaged over a field of view of approximately 4 mm. Informed consent was obtained from the subjects, experimental procedures adhered to the tenets of the Declaration of Helsinki, and study authorization was obtained from the appropriate local ethics review boards (clinical trial NCT04129021). Digital holograms resulting from the interference of light backscattered by the retina and an on-axis reference beam are recorded with a camera running at 60 to 75 kHz. The data processing applied to the reconstructed holograms consists of measuring the local optical temporal fluctuations of the holograms in order to reveal blood flow. In our previous work, this was done by calculating the temporal Fourier transform of the holograms over a short-time window of a few hundreds holograms. Then so-called power Doppler images were obtained by high-pass filtering of the DPSD. In order to implement a spatio-temporal filtering, an additional SVD filtering operation is here introduced prior to the calculation of the temporal Fourier transform. We used short-time windows of 1024 holograms, which gives a temporal resolution of 13.7 ms with a sampling frequency of 75 kHz. The subsequent data processing of the Doppler signal is left unchanged from our previous work~\cite{Puyo2018, Puyo2019, Puyo2019b}. For a given short-time window of $n_t$ consecutive holograms of spatial dimension $n_x \times n_y$, the stack of holograms is reshaped into a 2D space-time matrix $H$ of size ($n_x \times n_y$, $n_t$), where each column of the matrix contains all the hologram pixels at a given time. A SVD allows to write $H$ as the product of 3 matrices $U$, $\Delta$, and $V$ such that:
\begin{equation} \label{eq:eq_SVD}
H = U \Delta V^{*}
\end{equation}
where $^{*}$ stands for the conjugate transpose. The process is represented in Fig.~\ref{fig_SVD_Processing}(a). $\Delta$ is a rectangular and diagonal matrix of size ($n_x \times n_y$, $n_t$) that contains $n_t$ singular values. Each singular value $\lambda_i$ is associated to both a spatial eigenvector $U_i(x,y)$ (i.e. a 2D image), and a temporal eigenvector $V_i(t)$ (i.e. a 1D pattern). The spatial and temporal eigenvectors are contained in the columns of the orthonormal matrices $U$ and $V$ of sizes ($n_x \times n_y$, $n_x \times n_y$) and ($n_t$, $n_t$), respectively. $H$ can be written using these eigenvectors as:
\begin{equation} \label{eq:eq_SVD_SpaceTime}
H(x,y,t) = \sum_{i=1}^{n_t}  \lambda_i U_i(x,y) V^{*}_i(t)
\end{equation}

where $\lambda_i$ are the scalar eigenvalues, ordered by decreasing value. Thanks to the SVD, $H$ is written as the sum of $n_t$ independent terms, that are each determined by a specific spatial distribution of energy, and by a pattern of temporal variations, weighted by the corresponding singular value. The purpose of the SVD in this work is to constitute a basis where contributions modulated by the same pattern of temporal variations are regrouped in the same eigenvectors. It is assumed that retinal and corneal motion generate the same Doppler signal over the whole field of view, so their contributions are spatially very well correlated, and can thus be isolated in some singular values. That would lead to a better discrimination of the Doppler fluctuations originating of eye motion from those of blood flow than with the Fourier decomposition. The question is then how to identify the singular values associated to eye motion and blood flow. Because artifacts such as eye motion have a larger spatial coherence, it can be expected that their contribution lies in the singular values of highest energy. Conversely, blood flow induces Doppler shifts from the motion of individual red cells. This is a Doppler signal that comes from the uncorrelated motion of independent scatterers, which results in a low spatial coherence. Consequently, blood flow signal is distributed in more singular values of lower energy. Therefore, by truncating the first $n_c$ singular values from the matrix $\Delta$, it is possible to isolate blood flow from eye motion:
\begin{equation} \label{eq:eq_SVD_filter}
H_{\rm f}(x,y,t) = \sum_{i=n_c+1}^{n_t}  \lambda_i U_i(x,y) V_i(t)
\end{equation}
Thus $H_{\rm f}$ is the cube of holograms filtered from the singular contributions of highest energy. Afterwards, the beat frequency spectrum $S$ of the holograms filtered by SVD is analyzed pixel-wise by the conventional short-time Fourier transform:
\begin{equation} \label{eq:eq_PSD}
S(x,y,t_{n},f) = \left| \int_{t_{n}}^{t_{n}+ t_{\rm win}} H_{\rm f}(x,y,\tau) \exp{\left(-2 i\pi f \tau\right)} \, {\rm d}\tau \right|^2 
\end{equation}
where $t_{\rm win}$ denotes the extent of the short-time window, and $f$ is the temporal frequency. Finally, the bandpass filtered DPSD is integrated over a given frequency range $[f_1, \, f_2]$ to obtain power Doppler images $M_{0}(x,y,t_{n})$ that reveal blood flow:
\begin{equation}\label{eq:eq_PowerDoppler}
M_{0}(x,y,t_{n}) = \int_{f_1}^{f_2} S(x,y,t_{n},f) \;  df
\end{equation}
The process is then repeated for the next short-time window in order to obtain a time-resolved movie of blood flow fluctuations. In Fig.~\ref{fig_SVD_Processing} is shown an example of SVD on LDH data. The LDH measurement was acquired at 75 kHz in $n_x n_y = 512 \times 512$ pixels format, and $n_t=1024$ complex-valued holograms are used for the computation of the SVD. In Fig.~\ref{fig_SVD_Processing}(b) is plotted the energy of the $n_t$ ordered singular values in a logarithmic scale. The vertical red dashed line indicates a possible threshold $n_c = 34$ to separate clutter from blood flow. In Fig.~\ref{fig_SVD_Processing}(c) is shown the power spectrum density of the ordered temporal eigenvectors weighted by singular values: each column of the image displays in color the magnitude in dB of the DPSD normalized by its maximum value. The horizontal dashed line represents the filtering that would be obtained by Fourier thresholding of the DPSD, whereas the vertical dashed line shows the filtering that is obtained by truncating the singular value matrix. A few individual or averaged spatial eigenvectors are shown in Fig.~\ref{fig_SVD_Processing}(d), the notation $U_{m-n}$ represents the image obtained by averaging the eigenvectors over the range $m$ to $n$. As postulated the first singular values originate from sources of clutter whereas the rest of the singular values reveal blood flow. It is interesting to observe that singular values of lower energy tend to reveal faster blood flow: the spatial eigenvectors from 100 to 200 reveal smaller blood vessels than those from 200 to 500. What is crucial to observe in Fig.~\ref{fig_SVD_Processing}(c) however, is that the SVD isolates in the first eigenvectors (cf. arrow) some contributions of very high frequencies that only reveal clutter, as it is visible from the corresponding spatial eigenvectors. These unwanted contributions are at very high frequency so they cannot be filtered by high-pass temporal filtering. This is why the SVD provides a better basis than the Fourier space to filter clutter from blood flow. Setting aside these singularities, the eigenvectors corresponding to singular values of high energy tend to have slower variations whereas singular values of lower energy are associated to faster variations.

The choice of the threshold to truncate the singular value matrix is delicate. In the field of Doppler ultrasound, some techniques have been proposed based on the characteristics of the spatial or temporal eigenvectors~\cite{Baranger2018}. In this work we have used a fixed threshold based on the equivalent temporal frequency cutoff. For a window of $n_t$ holograms, whether with the SVD or fast Fourier transform (FFT), the cube of $n_t$ holograms is projected onto a set of $n_t$ vectors. Typically, for a measurement at $f_S = 60 \, \rm kHz$ and a short time window of $n_t = 1024$ holograms, a threshold equivalent to $f_1 = 2 \, \rm kHz$ is used: the SVD filtering is performed as described by setting $n_c$ to the integer nearest to $2.n_t . f_1  / f_S$. Then the power Doppler image is calculated with the $f_1$ lower cutoff frequency. This method amounts to rejecting the same number of elements with both decompositions. The choice of the 2 kHz threshold is the result of a compromise: a threshold too low comes with the risk of not rejecting all clutter, and a threshold too high prevents imaging slower blood flow.

Computationally, the short-time window analysis was performed sequentially (i.e. one window at a time) with a NVIDIA Titan RTX GPU card on single precision floating-point complex-valued arrays with MATLAB R2019b. For a window of size 512x512x1024, the numerical propagation takes about 0.5 s, and the fast Fourier transform and power Doppler summation about 0.1 s. The SVD basis is computed from the eigendecomposition of the hologram covariance matrix. The total time added by the SVD filtering is about 1.5 s. With the additional time to read the data from the hard disk drive and additional minor operations, one window is processed in about 6.5 s. The complete time required to process a stack of 259.329 holograms in 512x512 (about 3.5 s of acquisition with a 75 kHz camera frame rate, this is the maximum possible from the camera on-board memory) with short-time windows overlapping by half, is about an hour.

\section{Results}

\subsection{Results in different situations}

\begin{figure}[t!]
\centering
\includegraphics[width = 1\linewidth]{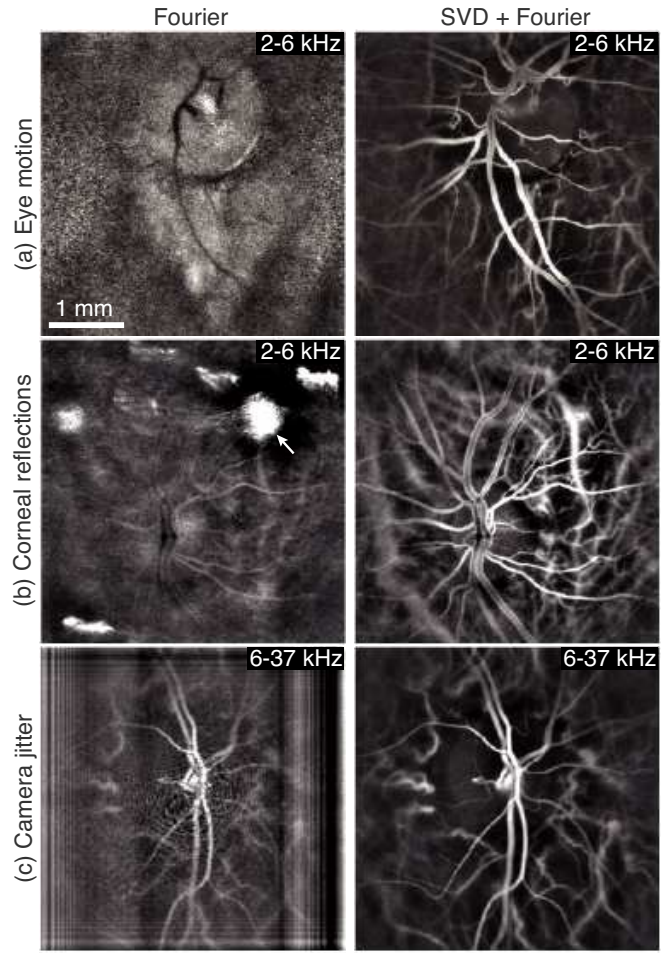}
\caption{Blood flow images improvement thanks to the SVD filtering. The left and right column show the standard power Doppler images and their SVD filtered version in the presence of: (a) retinal motion, (b) corneal reflections, and (c) camera jitter.
}
\label{fig_Images_WithWithoutSVD}
\end{figure}

We first demonstrate the efficiency of the SVD filtering in Fig.~\ref{fig_Images_WithWithoutSVD} on LDH measurements strongly affected by motion artifacts when using the standard highpass Fourier filter. The power Doppler images calculated without and with the additional SVD filter are shown on the left and right, respectively.
In Fig.~\ref{fig_Images_WithWithoutSVD}(a), the 2-6 kHz power Doppler image obtained just with the Fourier highpass is largely dominated by strong retinal motion, which prevents blood flow from being revealed. However the SVD filtering applied to the same data allows to reject the retinal motion, so that blood flow can be revealed on the power Doppler image. In Fig.~\ref{fig_Images_WithWithoutSVD}(b), the measurement was similarly affected by a strong corneal reflection (arrow) that could not be removed by Fourier filtering on the 2-6 kHz frequency range. Again, this spurious contribution can be efficiently removed with the SVD filtering. Finally, in the third example shown in Fig.~\ref{fig_Images_WithWithoutSVD}(c) is demonstrated a case where the power Doppler image was strongly affected by camera jitter. Unlike previous examples, this time the parasitic signal affects the power Doppler image on the frequency range 6-37 kHz. This unwanted contribution takes the form of a diffraction pattern from the camera jittering pixel. Once again the SVD filtering allows to remove this artifactual contribution from the power Doppler image so that a clean image of blood flow can be produced.

These examples demonstrate that a spatio-temporal filtering of the holograms is efficient to remove unwanted contributions characterized either by a strong spatial coherence in the case of eye motion, or by a specific temporal pattern in the case of camera jitter. Thanks to their higher spatial coherence, these clutter contributions are associated with singular values of higher energy, which are discarded by the SVD filter because it removes the first singular values. The SVD is able to provide a signal decomposition basis more adapted to the discrimination of clutter from blood flow than the Fourier basis. This way, the additional SVD filtering allows to access Doppler frequency ranges otherwise dominated by clutter. This digital filter is not only able to isolate the Doppler signal of blood flow from that of eye motion, but it is also able to compensate for flaws of the physical detection channel with a particular spatio-temporal pattern, such as the camera jitter.

\subsection{SVD filtering of the eye motion}

\begin{figure}[t!]
\centering
\includegraphics[width = 1\linewidth]{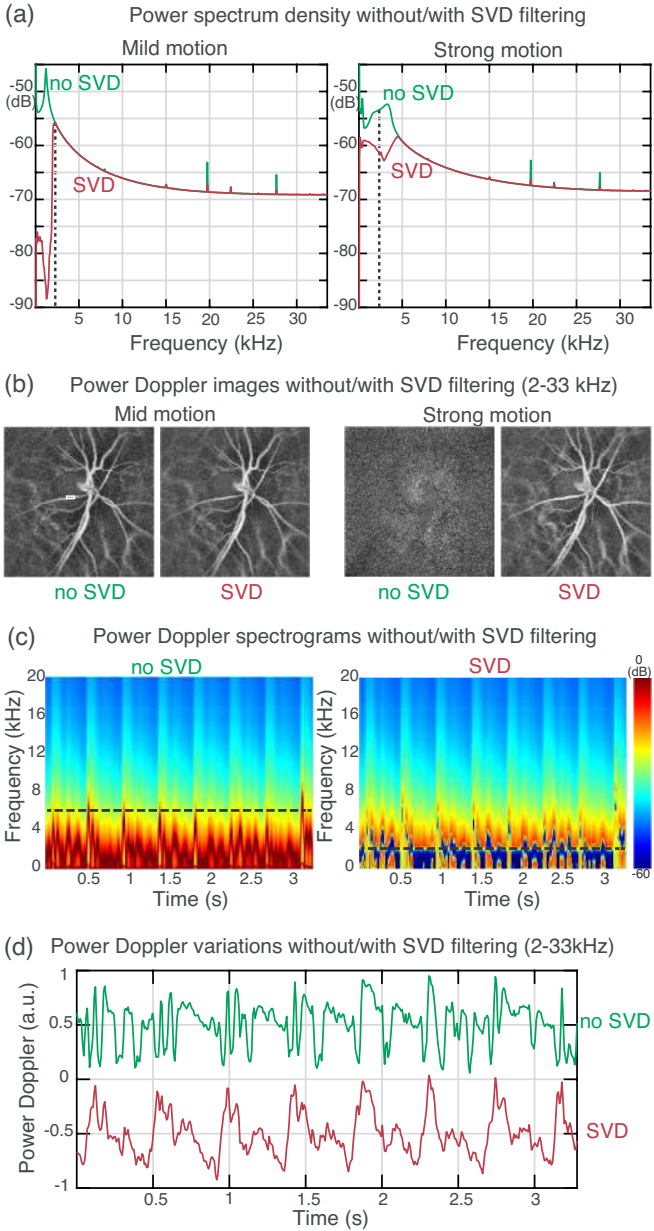}
\caption{Filtering the eye movements with SVD. (a) DPSD in the presence of mild and strong motion, the vertical dashed line indicates the 2 kHz threshold used to calculate the power Doppler. The SVD filtering removes contribution of higher frequencies than the 2 kHz cutoff. (b) Corresponding power Doppler images: in the presence of strong motion the SVD allows to preserve the blood flow signal.(c) Spectrograms ($t_{\rm win} =  15.3 \; \rm ms$): the SVD filter rejects the pulsatile eye motion thus allowing to lower the frequency cutoff. (d) Power Doppler variations in the artery ($t_{\rm win} =  15.3 \; \rm ms$). 
}
\label{fig_PlotPulsatility_Spectrum}
\end{figure}

During the few seconds of a LDH measurement can occur fixational eye movements~\cite{Martinez2006}, and cardiac related axial retinal and corneal motion~\cite{Singh2010}. In Fig.~\ref{fig_PlotPulsatility_Spectrum}, we show the effect of the SVD filtering in the presence of intermittent eye motion. An acquisition performed at 66 kHz is considered, and the 2-33 kHz frequency band is used to reveal retinal blood flow for all of the displayed power Doppler images and power Doppler variations in the Fig. The short-time window is positioned at two moments corresponding to cases of mild and strong eye motion, where mild and strong motion are defined as generating Doppler signals below and above the 2 kHz threshold, respectively. All the plots displayed in the Fig. are measured in a region of interest positioned on a small artery as shown in Fig.~\ref{fig_PlotPulsatility_Spectrum}(b).

We first investigate the effect of the SVD filtering on the DPSD. In Fig.~\ref{fig_PlotPulsatility_Spectrum}(a) are represented the power spectra density of $H$ and $H_f$ in the presence of mild (left) and strong (right) eye motion. The vertical dashed line indicates the 2 kHz cutoff frequency used to calculate the power Doppler. In the presence of mild motion, the DPSD with and without prior SVD filtering are essentially the same above the cutoff frequency: the SVD filtering only rejects signals of frequencies lower than the cutoff. As the power Doppler is calculated as the integral of the spectrum above the cutoff frequency, in the presence of mild motion the power Doppler images with and without SVD filtering shown below in Fig.~\ref{fig_PlotPulsatility_Spectrum}(b) are close to identical. In the presence of strong eye motion however, the spectra now differ significantly above the cutoff frequency. Without SVD filtering, the DPSD shows a peak around 3 kHz corresponding to the eye movement. This contribution is integrated in the power Doppler, and the power Doppler image without SVD filtering is dominated by eye motion. For its part, the SVD filtered DPSD exhibits a dip around the same frequency where there should be the eye motion peak because the eye motion has been rejected. The corresponding power Doppler image is now able to reveal blood flow.

The spectrograms of the holograms fluctuations with and without SVD filtering shown in Fig.~\ref{fig_PlotPulsatility_Spectrum}(c) demonstrate how the SVD filter suppresses the power Doppler contribution due to eye motion occurring during cardiac cycles. The magnitude of the DPSDs for all short-time windows of the movie are represented in the column of the spectrogram in color tones. The dB values are obtained by calculating the decimal logarithm of the DPSD normalized by its maximum value. On the spectrogram on the left, showing the DPSD without prior SVD filtering, the eye motion occurring throughout cardiac cycles can be identified in red in the 0-6 frequency range. The frequency threshold that would be required to filter eye motion is represented by the horizontal dashed line, at approximately 6 kHz. Below this 6 kHz the signal contains strong eye motion contribution, and above is essentially fast blood flow. The spectrogram on the right shows the DPSD of the data filtered by SVD. It is possible to see that whenever it occurs, global tissue motion is effectively rejected. This allows to lower the frequency threshold used for the calculation of the power Doppler down to 2 kHz (horizontal dashed line). The SVD filter allows to access the 2-6 kHz range without being so much affected by eye motion artifacts. It is also possible to observe that when there is no eye movement, the SVD only removes the signals below 2 kHz, because the SVD and the Fourier space provide very similar decompositions. This effect is also visible in Fig.~\ref{fig_SVD_Processing}(c) from the linear aspect of the Fourier decomposition of the ordered temporal eigenvectors.


Finally, the robustness with which the SVD filtering is able to reject eye motion as it occurs throughout cardiac cycles is demonstrated in Fig.~\ref{fig_PlotPulsatility_Spectrum}(d). Power Doppler variations are measured in the small artery on the 2-33 kHz range, again with and without SVD filtering. These two plots show that without SVD filtering, the blood flow waveform cannot be revealed because there are too many eye movements. However when using the SVD filtering, the blood flow waveform can be revealed as eye motion contribution is effectively canceled. The two corresponding movies showing the variations of power Doppler without and with prior SVD filtering are shown on the left and right, respectively, in \textcolor{blue}{\href{https://youtu.be/RZD6coK74MQ}{Visualization 1}}.

What can be concluded from Fig.~\ref{fig_PlotPulsatility_Spectrum} is that in the presence of mild motion the SVD filtering does not modify the power Doppler image as the SVD and Fourier signal decomposition bases do not differ significantly. However in the presence of strong motion, there is a contribution of high spatial coherence that leads to a difference between the SVD and Fourier basis. Then the SVD is able to reject those contributions that are of higher frequencies than the power Doppler cutoff frequency of the measurement band, and thus dramatically improves how power Doppler images reveal blood flow. The SVD filtering opens the possibility to exploit the low frequency ranges without being hindered by eye motion, i.e. LDH can measure lower flow velocities over cardiac cycles. This allows for example to combine low and high frequency ranges in composite images as detailed in~\cite{Puyo2019} in blood flow movies, instead of doing it only on averaged images as before; this is shown in \textcolor{blue}{\href{https://youtu.be/6BZpJ87LeNk}{Visualization 2}}.

\subsection{Retinal blood flow in pathology}

\begin{figure}[t!]
\centering
\includegraphics[width = 1\linewidth]{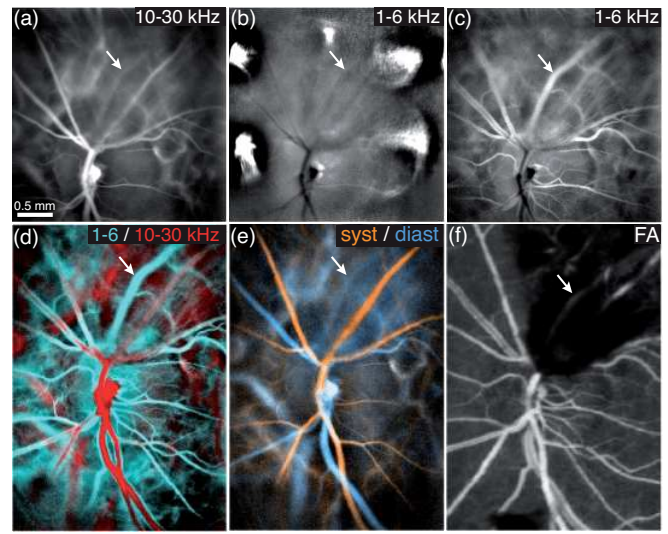}
\caption{Imaging blood flow in a case of retinal vein occlusion. (a) The 10-30 kHz power Doppler image does not show blood flow in the occluded vein. (b) the 1-6 kHz range fails to reveal it because of eye motion. (c) With a SVD filtering the blood flow at low frequency can be revealed. (d) Composite image of the low/high flow in cyan/red; the occluded vessel (arrow) can be easily identified. (e) Composite image of the systole/diastole in orange/blue. (f) Fluorescein angiography in the same region cannot image through the preretinal hemorrhage.
}
\label{fig_CasClinique}
\end{figure}

Many retinal pathologies are associated with a decreased blood flow, which is challenging to image with LDH as it requires to work with lower frequencies. In Fig.~\ref{fig_CasClinique}, we show an example of blood flow imaging in a case of retinal vein occlusion. The superior hemivein is occluded, and preretinal hemorrhage has formed above it. As shown in Fig.~\ref{fig_CasClinique}(a), the 10-30 kHz power Doppler reveals all the large retinal vessels except the occluded vein because its blood flow is reduced. Without SVD filtering, the 1-6 kHz power Doppler image in Fig.~\ref{fig_CasClinique}(b) also fails to reveal the occluded vein because there are too many artifacts. However when applying the SVD filter in Fig.~\ref{fig_CasClinique}(c), the artifacts can be removed, and the 1-6 kHz power Doppler image is now able to reveal the flow in the occluded vein.

In Fig.~\ref{fig_CasClinique}(d) and Fig.~\ref{fig_CasClinique}(e) are displayed composite images obtained by stitching two LDH measurements. In Fig.~\ref{fig_CasClinique}(d), the 1-6 and 10-30 kHz frequency ranges are fused into a composite image that shows low flow in cyan and high flow in red. This image is very efficient to evidence the blood flow disruption in the superior hemivessels: healthy vessels of that size would normally show, like the inferior hemivessels, a red lumen surrounded by an outer cyan sheath. However the occluded vein appears only in cyan, and the artery feeding the occluded vascular territory appears in red, with a weaker contrast. The image shown in Fig.~\ref{fig_CasClinique}(e) was obtained by fusing two images averaged from the high frequency power Doppler movie during arbitrarily chosen periods corresponding to systole and diastole. This image allow to easily identify arteries in orange and veins in blue thanks to their particular flow variation patterns. The image obtained in the same region by fluorescein angiography shown in Fig.~\ref{fig_CasClinique}(f) is not able to reveal the blood vessels through the preretinal hemorrhage.

This case of retinal vein occlusion shows that the SVD filter is necessary to detect an abnormal perfusion, because in a situation where blood flow is reduced the signal is revealed by the low frequencies subject to many parasitic Doppler contributions. It also demonstrates how LDH is relevant to study non-invasively retinal vascular diseases thanks to its ability to assess blood flow over an extended field of view. Furthermore, considering the specific dynamic fluctuations of blood flow that exist in cases of retinal vein occlusions~\cite{Paques2005}, LDH could prove particularly useful for the study of this pathology.

\subsection{Imaging the choroid}

\begin{figure}[t!]
\centering
\includegraphics[width = 1\linewidth]{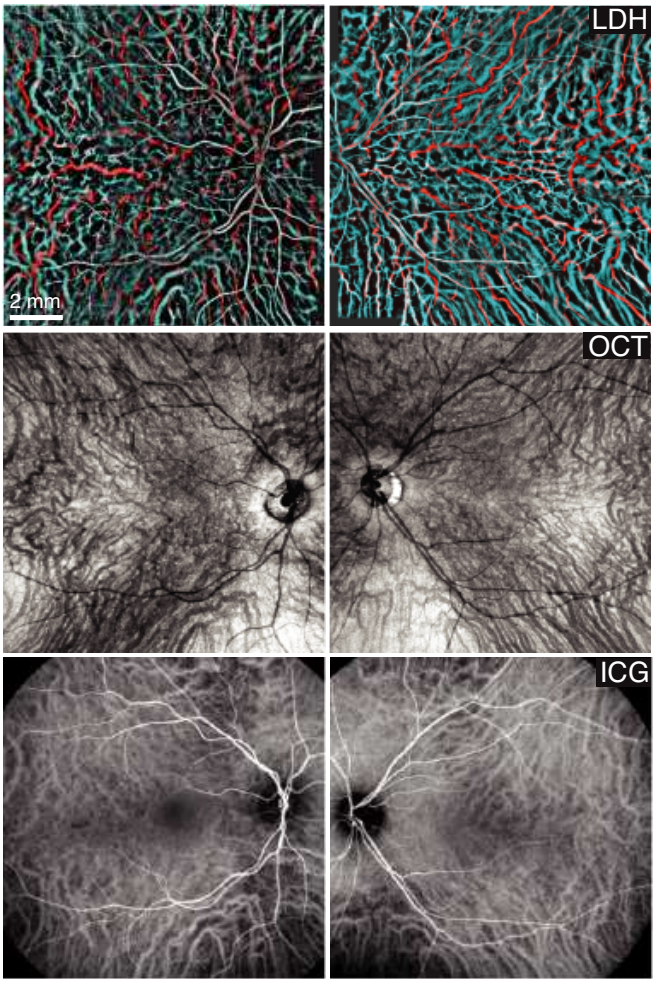}
\caption{Imaging the choroid with LDH, OCT (Plex Elite 9000, Zeiss), and ICG-angiography (Spectralis, Heidelberg). The LDH montages have been stitched from 5x5 images obtained by fusing the 2-6 and 10-30 kHz frequency ranges in cyan and red. The access to the low frequency range made routinely possible by SVD is critical to reveal choroidal veins.
}
\label{fig_ChoroidPanorama}
\end{figure}

We previously showed that there are discrepancies of flow velocities between arteries and veins in the choroid which can be used to perform an arteriovenous differentiation~\cite{Puyo2019}. This was previously demonstrated on a limited number of power Doppler images as it required manual filtering of images affected by motion artifacts. The SVD filtering is robust enough so that the low frequency range used to reveal the choroidal veins can now be accessed systematically, and manual filtering is no longer necessary. In the first row of Fig.~\ref{fig_ChoroidPanorama}, we show two LDH montages revealing the choroid obtained by stitching 5x5 individual images. The field of view on the retina of these montages is estimated to be approximately 12 mm large. Thanks to the SVD filter, the making of the 50 individual composite power Doppler images could be fully automated, and the low frequency range could be accessed in all measurements. These two LDH montages reveal efficiently all the choroidal vessels, and the functional blood flow contrast offered by LDH allows to identify the vessels type. In the second and third rows of Fig.~\ref{fig_ChoroidPanorama}, we show the images obtained in the same eyes with optical coherence tomography (OCT) and indocyanin green angiography (ICG), which are the standard imaging modalities currently available in clinics to study the choroid en-face. Both OCT and ICG are not as efficient as LDH to reveal the choroidal arteries: the blood flow contrast brought by LDH allows to have a better understanding of the choroid anatomy. As shown earlier in this article, the SVD filtering is critical to access the low frequency range (here 2-6 kHz), which is the frequency range that reveals the choroidal veins and the smaller arteriolar vasculature.

\section{Conclusion}
The access to low frequencies in laser Doppler imaging is critical to improve the sensitivity to slower blood flow. Low frequency shifts are needed not only to reveal blood flow in smaller vessels, but also in large vessels where blood velocity is reduced because of vascular pathologies such as occlusions. It is also important to be able to use the low frequencies when imaging the choroid because veins are revealed by lower frequency shifts due to the lower velocities of flow in these vessels. Unfortunately, low Doppler frequency shifts from slow blood flow overlap in the Fourier space with the Doppler response of unwanted contributions such as retinal and corneal motion. Adding the spatial dimensions into the analysis of the holograms fluctuations allows to take advantage of the greater spatio-temporal coherence of eye motion compared to the uncorrelated Doppler shifts of individual red cells. We have shown that rejecting the singularities of highest energy is a simple way to filter clutter from blood flow and greatly improve the quality of low frequency power Doppler images. Additionally, because the SVD generates a basis adapted to each short-time window, it is able to filter pulsatile or fixational eye movement during cardiac cycles. This dramatically improves the quality of the blood flow waveform measured at low frequency, and is very valuable to image blood flow in the eye of subjects with limited fixation ability. Finally, the SVD filter is also able to compensate for the camera jitter, thereby reducing the camera requirements. Overall, the SVD filtering allows to significantly lower the frequency threshold used to calculate the power Doppler and detect slower blood flow more robustly.
This spatio-temporal analysis of the holograms fluctuations is effective because measurements are performed in full-field. Indeed it is necessary that all pixels receive the same signal so that it is correlated over the field of view and becomes a singularity in the SVD. Moreover, it seems advantageous to dispose of a dense sampling of the interferograms intensity variations both in time and space, as provided by a high-speed camera, so that the SVD can more efficiently identify patterns of temporal variations with strong spatial coherence. The spatio-temporal filtering demonstrated in this article confers LDH a decisive advantage compared to scanning laser Doppler techniques.

In conclusion, a spatio-temporal filtering can dramatically improve the LDH signal quality at low frequency in various situations. By doing so, it addresses the key issue of detecting slower blood flow, which is critical to reveal smaller vessels, vessels with a pathologically reduced blood flow, and choroidal veins. The ability to reveal blood flow in smaller retinal vessels will be especially helpful to LDH to better reveal blood vessels in the optic nerve head, which could be of high clinical interest for the study of optic neuropathies. LDH appears as a promising method to study the ocular vasculature in healthy and diseased eyes.

\section*{Funding Information}
This work was supported by the European Research Council (ERC Synergy HELMHOLTZ, grant agreement \#610110). The Titan RTX graphics card used for this research was donated by the NVIDIA Corporation.

\section*{Acknowledgments}
The authors would like to thank J\'er\^ome Baranger for helpful discussion, and the OVR association.

\section*{Disclosures}
The authors declare no conflicts of interest.

\section*{Supplementary Material}
\noindent
\textcolor{blue}{\href{https://youtu.be/g4zVGLlw9yw}{Supplementary Visualization 1}}. \newline
\textcolor{blue}{\href{https://youtu.be/AA9Q_2B9FZA}{Supplementary Visualization 2}}. \newline
\textcolor{blue}{\href{https://youtu.be/wlQbjTSmpKk}{Supplementary Visualization 3}}. \newline

\bibliography{../Biblio/Bibliography}

\end{document}